\documentclass[preprint,10pt, letterpaper]{elsarticle}

\usepackage{amssymb}
\usepackage{hyperref}
\usepackage{booktabs}
\usepackage{array}
\usepackage{float}
\usepackage{pdfcomment}
\usepackage{amsmath}
\usepackage{cancel}
\usepackage{graphicx}
\usepackage{subcaption}
\usepackage{makecell}
\usepackage{xcolor}
\usepackage{multicol}
\usepackage{listings}
\usepackage{algorithm}
\usepackage{algpseudocode}
\usepackage{caption}
\usepackage{enumitem}

\usepackage[margin=1.0in]{geometry}

\journal{SoftwareX}

\begin{document}
\renewcommand{\labelenumii}{\arabic{enumi}.\arabic{enumii}}

\begin{frontmatter}

\title{\textsc{CodeJeNN}: A simple C\texttt{++} neural network generator for physics applications}

\author[SDSU]{Jay Arcities}
\author[SDSU]{Pavel Popov}
\author[NRL]{Eric J. Ching}
\author[NRL]{Kamal Viswanath}
\author[NRL]{Ryan F. Johnson}

\address[SDSU]{San Diego State University, Aerospace Engineering Department, San Diego, CA 92182}
\address[NRL]{U.S. Naval Research Laboratory, Laboratories of Computational Physics and Fluid Dynamics, Washington, DC 20375}

\begin{abstract}
Machine learning has shown promise for physics applications, but integrating Python-based libraries into high-performance C\texttt{++} solvers can create performance bottlenecks. We present \textsc{CodeJeNN}, which addresses this gap by auto-generating self-contained C\texttt{++} code from trained Keras models for inference. This eliminates external dependencies through inlined functions, allowing seamless integration into existing frameworks. We describe the Keras-to-C\texttt{++} workflow, supported architectures, and limitations. \textsc{CodeJeNN} is demonstrated through inference benchmarks against Keras (eager and optimized) and a computational fluid dynamics (CFD) simulation of a hydrogen-air mixing layer in which viscosity is modeled using a neural network, showing that the generated models integrate directly into a solver while remaining efficient and accurate.
\end{abstract}


\end{frontmatter}


\begin{table}[!h]
    \begin{tabular}{|l|p{6.5cm}|p{6.5cm}|}
        \hline
        \textbf{Nr.} & \textbf{Code metadata description} & \textbf{Please fill in this column} \\
        \hline
        C1 & Current code version & v1.0 \\
        \hline
        C2 & Permanent GitHub link to code/repository used for this code version & \url{https://github.com/jarcities/codejenn} \\
        \hline
        C3 & Legal Code License & NRL Open License Agreement \\
        \hline
        C4 & Code versioning system used & Git \\
        \hline
        C5 & Software code languages, tools, and services used & Python 3.12, C\texttt{++}23 \\
        \hline
        C6 & Compilation requirements, operating environments \& dependencies & Keras 3, Anaconda Distribution, further description below \\
        \hline
        C7 & If available Link to developer documentation/manual & \url{https://github.com/jarcities/codejenn/blob/main/readme.md} \\
        \hline
        C8 & Support email for questions & jarcities@sdsu.edu \\
        \hline
    \end{tabular}
    \caption{Code metadata}
\end{table}

\let\svthefootnote\thefootnote\let\thefootnote\relax\footnotetext{\\ \hspace*{25pt}DISTRIBUTION STATEMENT A. Approved for public release. Distribution is unlimited.}\addtocounter{footnote}{-1}\let\thefootnote\svthefootnote

\section{Motivation and Significance}
Machine learning has shown promise when integrated with numerical schemes for partial differential equations (PDEs) as well as physical modeling, enabling research in data-driven preconditioners \cite{hausner_learning_2024, li_learning_2023}, surrogate constitutive models \cite{fuhg_review_2025}, and physics-informed neural networks \cite{raissi_physics-informed_2019, zhao_comprehensive_2024, cai_physics-informed_2021}. In computational fluid dynamics (CFD), neural networks are applied to accelerate simulations, turbulence modeling, and reduced order models \cite{vinuesa_enhancing_2022}. Modern neural network architectures are accessible through high-level libraries such as \textsc{ONNX} \cite{bai_onnx_2019}, TensorFlow \cite{martin_abadi_tensorflow_2015}, Keras \cite{chollet_keras_2015}, and \textsc{PyTorch} \cite{ansel_pytorch_2024}.
However, embedding these models into a physics solver is not straightforward. Most physics solvers are written in compiled languages (e.g., Fortran and C\texttt{++}) for better resource control, while Machine Learning (ML) libraries are primarily Python-based. This disparity often adds computational overhead. Furthermore, integrating complex ML architectures into established codes often requires significant implementation effort and experimentation.

The gap between ML libraries and compiled solvers often forces researchers to rely on framework-provided C\texttt{++} APIs, which can be incompatible with established parallelization and vectorization codes. Delegating tensor operations to external APIs limits optimization and adds dependencies that complicate integration.
For example, Partee et. al. \cite{partee_using_2022} developed SmartSim, which uses an in-memory client-server database (Redis) to decouple numerical solvers from deep learning frameworks, allowing for offloading of runtime inference to GPUs with minimal runtime overhead.
Bati and Bryngelson \cite{bati_rosenna_2024} developed \textsc{RoseNNa}, a \textsc{Fortran}-based encoder that generates executable code for Multi-Layer Perceptrons (MLP) and Recurrent Neural Nets (RNN) from \textsc{ONNX} models.
While \textsc{ONNX} excels as a model exchange format across machine learning frameworks, Keras is a high-level deep learning API that abstracts away low-level tensor operations to provide a user-friendly environment for rapidly training and deploying neural networks. To directly bridge these user-friendly Keras workflows with modern C\texttt{++} solvers, we have developed \textsc{CodeJeNN}.
\textsc{CodeJeNN} produces all supporting activation and propagation step functions in plain C\texttt{++}, allowing users to optimize source code directly without external dependencies such as \textsc{Lapack} or \textsc{Eigen}.

\textsc{CodeJeNN} is designed to be extensible. Each layer type is implemented as an independent module. Supporting a new architecture therefore follows the same steps every time. An LLM generates only the logic for the missing layer, without changing the implementation of the existing ones. This provides a non-invasive approach to extending the framework while minimizing token consumption and reducing the effort required to experiment with new architectures.

The remainder of the paper is structured as follows: Section 2 covers neural network principles; Section 3 describes the \textsc{CodeJeNN} architecture and compatibility; Section 4 demonstrates the software through inference speed benchmarks and a CFD viscosity modeling case using Wilke's model \cite{wilke_viscosity_1950}; and Section 5 discusses the broader impact of \textsc{CodeJeNN}, including its extensibility, potential applications, and future development.

\section{Machine Learning Background}
\begin{figure}[h!]
    \centering
    \includegraphics[width=0.45\linewidth]{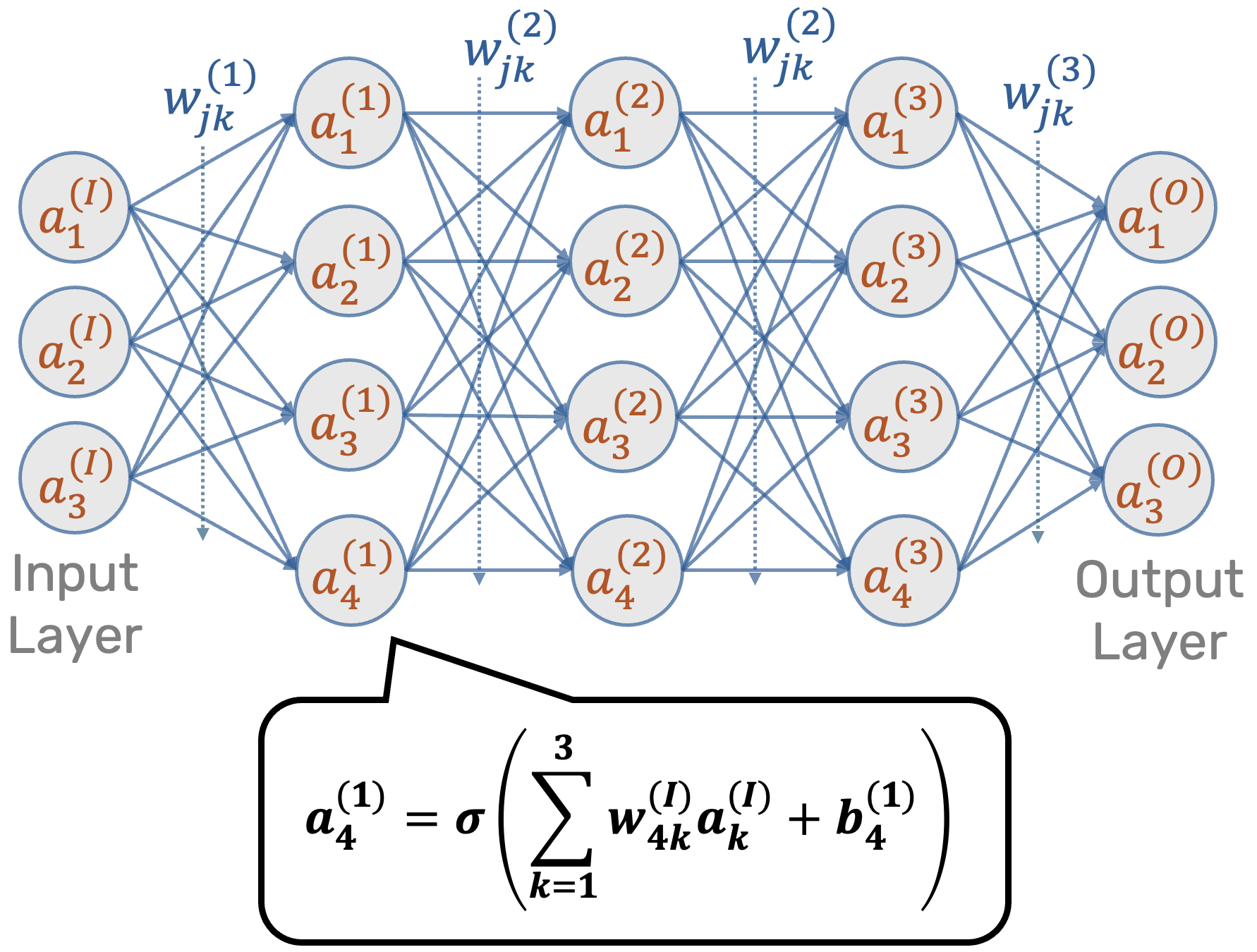}
    \caption{Schematic of a feedfoward network (MLP), where $a_j^{(i)}$ is the output from a given neuron, $w_{jk}^{(i)}$ is the specific weight sent to that neuron given the edge. $I$ and $O$ in $a_j^{(i)}$ denote the input and output neurons. The superscript indexes the layers, and the subscript indexes the neurons of a given layer. The expression for $a_4^{(1)}$ is shown in the message box.}
    \label{fig:mlp_schematic}
\end{figure}

A neural network is a surrogate model used to approximate complex mappings from training data. \textsc{CodeJeNN} generates code for MLPs and Convolutional Neural Networks (CNN). In an MLP, each layer performs an affine transformation using weights $\mathbf{w}$ and biases $\mathbf{b}$, followed by an activation function $\mathbf{\sigma}$ to introduce non-linearity, as
\begin{gather}\label{eq_affine_trans}
    \mathbf{a} = \mathbf{\sigma}\left(\mathbf{w}^T \mathbf{x} + \mathbf{b}\right).
\end{gather}
A typical MLP schematic is shown in Figure \ref{fig:mlp_schematic}.

\begin{figure}[h!]
    \centering
    \includegraphics[width=\linewidth]{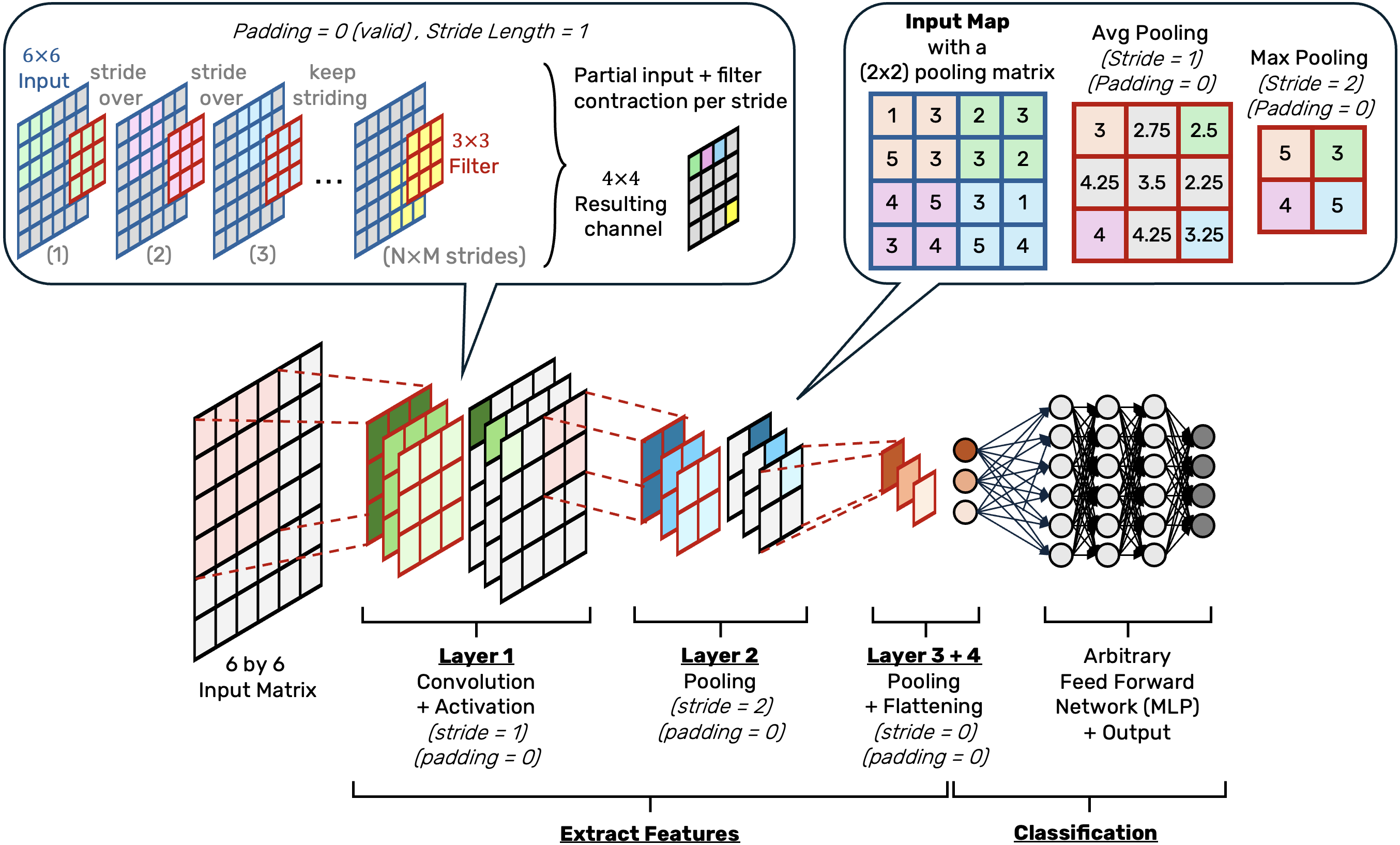}
    \caption{Schematic of a CNN with one convolution layer and two pooling layers. The kernels in the first two layers have three filters. The MLP after layer four is of arbitrary shape and used for classification. The operation for applying a $3\times3$ convolution kernel made of three filters is shown in the top left message box. The operation of applying a $2\times2$ pooling matrix for both average and max pooling is shown in the top right message box. The resulting matrix of these operations is called a channel. The first layer has three channels.}
    \label{fig:cnn_schematic}
\end{figure}

CNNs extend this formulation using kernels to extract spatial features from multi-channel inputs:
\begin{gather}
    Y_{c_\text{out}, i, j} =
    \sum_{c_\text{in}}
    \sum_{m}
    \sum_{n}
    K_{c_\text{out}, c_\text{in}, m, n} \,
    X_{c_\text{in}, i+m, j+n}
    + b_{c_\text{out}},
    \label{eq_conv_filter_eq}
\end{gather}
Here, $c_\text{out}$ and $c_\text{in}$ are the input and output channel index of the filter, $m$ and $n$ are the spatial indices in the filter, $i$ and $j$ are the indices of the input data, and $K$ denotes the filter. In Eq. \eqref{eq_conv_filter_eq}, the CNN layer output $Y_{c_{\text{out}}, i, j}$ is computed as a weighted combination of local input regions across all input channels, producing feature maps that capture local spatial patterns. A channel is the resulting map/matrix/image of a filter or pooling operation. A schematic of a CNN with one convolutional kernel (three filters) and two pooling layers is shown in Figure \ref{fig:cnn_schematic}. Comparing Figure~\ref{fig:mlp_schematic} and Figure~\ref{fig:cnn_schematic} shows that MLPs and CNNs differ in the layer types they use, the shape of the data passed between layers, and the parameters each layer requires. Each of these differences must be handled explicitly when a trained network is implemented in an existing codebase.

\section{Software Description}
\begin{figure}[h!]
    \centering
    \includegraphics[width=.5\linewidth]{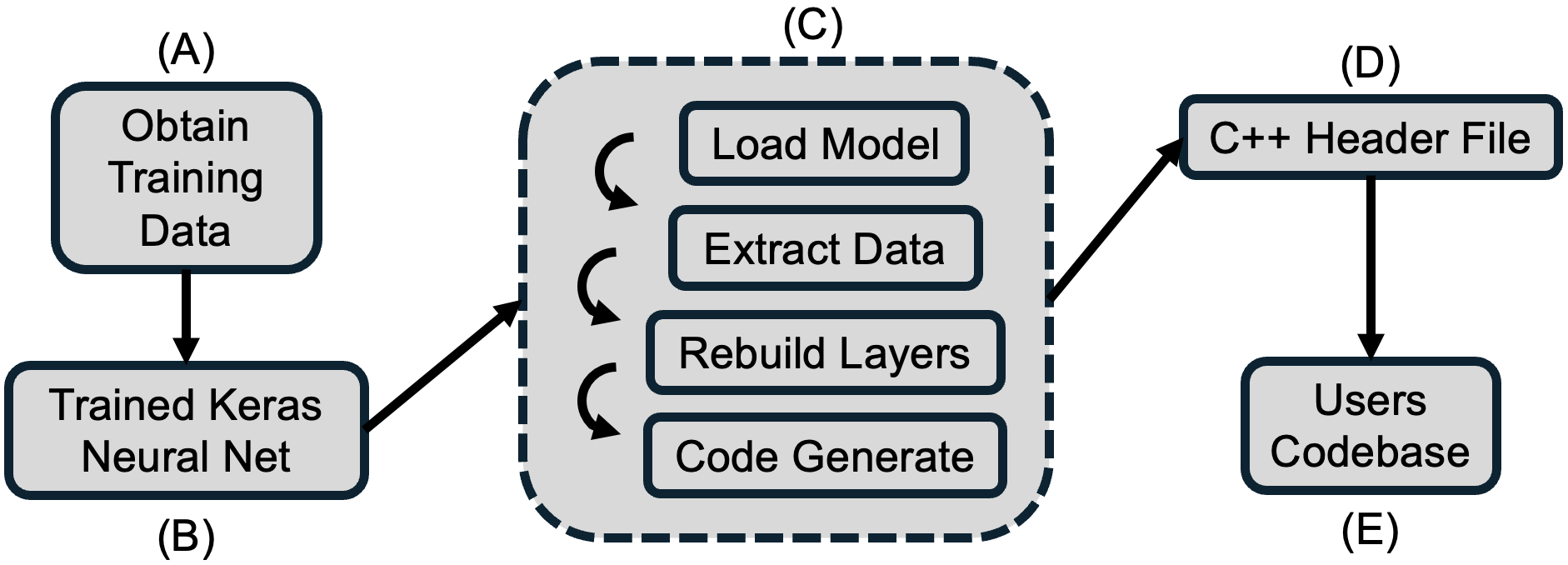}
    \caption{\textsc{CodeJeNN} (C) code generates a \textsc{Keras}-trained neural network into a C\texttt{++} header file for the user's computational physics solver (E).}
    \label{fig:codejenn_flowchart}
\end{figure}

Figure \ref{fig:codejenn_flowchart} illustrates the workflow of \textsc{CodeJeNN}. The process begins with the user obtaining data to train a neural network. That data is used to train a model using Keras, and the parameters and hyperparameters are saved in a .keras or .h5 file. The user then utilizes \textsc{CodeJeNN} to read the file and generate self-contained C\texttt{++} code that reproduces the trained neural network. The main principle of CodeJeNN is to minimize external dependencies, as well as inline as much as possible, execute efficiently on local hardware, and remain fully self-contained using internal implementations.

\subsection{Keras Compatibility}
Keras is chosen for its high-level API and backend integration with major frameworks (e.g., JAX, TensorFlow, PyTorch). It stores the model architecture, hyperparameters, and parameters in a single file, allowing simplified model exchange and downstream \textsc{CodeJeNN} code generation. Its emphasis on readability and accessibility is important for researchers who wish to train models and automatically generate equivalent C\texttt{++} code without having to develop infrastructure in their codebase for model inference.

\subsection{Code Generation}
\textsc{CodeJeNN}'s code generation proceeds in four steps: (1) load the saved model file using Keras' \verb|load_model| function, including any custom activation functions; (2) extract model parameters into a Python dictionary, where users can extend support for additional layers; (3) reconstruct the model from input to output layer; and (4) convert to a C\texttt{++} header file.

Algorithm \ref{alg:mlp_structure_example} presents a step-by-step procedure for how \textsc{CodeJeNN} formats a generated MLP once the model is loaded in Python. First, the propagation step functions (similar to the neuron edges shown in Figure \ref{fig:mlp_schematic}) are defined before the MLP function header. Then, within the MLP function body, there are three main sections. First, each layer’s parameters are generated. Second, the activation functions are defined as C\texttt{++} lambda functions. Third, the function calls associated with the propagation step, where inference is performed, are produced. \textsc{CodeJeNN} generates neural networks in this format to enable downstream code optimization for users, where the functions associated with the propagation step benefit the most from modifications.

\begin{algorithm}[h]
    \caption{Example of a code-generated MLP}
    \begingroup
        \makeatletter
        \renewcommand{\alglinenumber}[1]{\makebox[1.5em][r]{\scriptsize #1}\hspace{1em}}
        \makeatother

        \begin{algorithmic}[1]
            \vspace{0.5\baselineskip}
            \State \textbf{Import packages:} iostream, array, cmath, functional, etc.

            \vspace{0.5em}
            \State \textbf{Define propagation step functions:}
            \State \hspace{1em} Dense(outputs, inputs, weights, biases, input size, activation)
            \State \hspace{1em} UnitNormalization(outputs, inputs, epsilon)

            \vspace{0.5em}
            \Function{MLP}{input}
                \State \textbf{Print model parameters:}
                \State \hspace{1em} Layer weights and biases
                \State \hspace{1em} Filter parameters
                \State \hspace{1em} Input/Output normalization constants

                \vspace{0.5em}
                \State \textbf{Define lambda activation functions:}
                \State \hspace{1em} linear(input)
                \State \hspace{1em} gelu(input)
                \State \hspace{1em} custom\_activation(input, index)

                \vspace{0.5em}
                \State \textbf{Propagation step:}
                \State \hspace{1em} Normalize $x \leftarrow (x - \mu_\text{in}) / \sigma_\text{in}$
                \State \hspace{1em} $h_1 \leftarrow \text{Dense}(x, W_1, b_1, linear)$
                \State \hspace{1em} $h_2 \leftarrow \text{UnitNormalization}(h_1, \epsilon)$
                \State \hspace{1em} $h_3 \leftarrow \text{gelu}(h_2)$
                \State \hspace{1em} $h_4 \leftarrow \text{Dense}(h_3, W_4, b_4, linear)$
                \State \hspace{1em} $h_5 \leftarrow \text{custom\_activation}(h_4)$
                \State \hspace{1em} Denormalize: $y \leftarrow h_5 \cdot \sigma_\text{out} + \mu_\text{out}$
                \State \Return $y$
            \EndFunction
            \vspace{0.5\baselineskip}
        \end{algorithmic}
    \endgroup
    \label{alg:mlp_structure_example}
\end{algorithm}

\subsection{Algorithm Design Structure}
\label{sec:alg_struct_specs}
The generated code is written such that all model parameters (weights, biases, and normalization constants) and layer dimensions are known at compile time, which facilitates incorporation into existing solvers and allows efficient cache utilization and fast inlined computation. As a result, an important design choice is the explicit use of syntax-level specifications, such as \texttt{void}, \texttt{inline}, \texttt{constexpr}, and \texttt{static}, which are applied consistently across arrays, headers, and function templates.
Although straightforward integration is the top priority, \textsc{CodeJeNN} still inlines generated code wherever possible so that the embedded model remains computationally efficient. Inlining code eliminates function-call overhead, improves CPU instruction caching, and allows compilers to aggressively optimize wherever necessary. Beyond performance, it generates self-contained, dependency-free code that helps avoid integration issues.

Most computational physics codes are already highly optimized, and further aggressive optimization of the generated code from \textsc{CodeJeNN} prior to integration can be counterproductive. For example, nested parallelism arising from both the neural network and its evaluation can lead to thread oversubscription, where more threads are launched than available \cite{yan_proposal_2016}, and thus degraded performance. As a result, compile-time optimizations and the use of external libraries are intentionally deferred to the user. The objective is to achieve readable and accessible neural networks that can be manipulated further once their implementation is satisfactory. This design choice also motivates placing functions corresponding to the propagation step at the top of the generated models (Algorithm~\ref{alg:mlp_structure_example}), where they are most amenable to user-directed optimization. In addition, loops within lambda-based activation functions and lower-level function calls are structured to allow straightforward parallelization or vectorization when appropriate.

\subsection{Capabilities and Limitations}
\textsc{CodeJeNN} supports the following activation functions:

\begin{multicols}{4}
    \raggedcolumns
    \begin{itemize}[itemsep=-5pt, topsep=0pt, leftmargin=*]
        \item[] \textsc{ReLU} \cite{nair_rectified_2010}
        \item[] \textsc{Sigmoid}
        \item[] \textsc{Tanh}
        \item[] \textsc{LeakyReLU} \cite{he_delving_2015}
        \item[] \textsc{Linear}
        \item[] \textsc{ELU} \cite{clevert_fast_2016}
        \item[] \textsc{Swish} \cite{ramachandran_searching_2017}
        \item[] \textsc{SiLU} \cite{ramachandran_searching_2017}
        \item[] \textsc{GeLU} \cite{hendrycks_gaussian_2023}
        \item[] \textsc{Softmax}
        \item[] \textsc{Mish} \cite{misra_mish_2020}
        \item[] \textsc{Softplus} \cite{dugas_incorporating_2000}
        \item[] \textsc{SeLU} \cite{klambauer_self-normalizing_2017}
        \item[] \textsc{Exponential}
        \item[] \textsc{CeLU} \cite{barron_continuously_2017}
        \item[] \textsc{PreLU}
    \end{itemize}
\end{multicols}

\noindent
\textsc{CodeJeNN} also supports custom activation functions; examples are provided in the tutorials. Normalization parameters for inputs and outputs are also supported.

\textsc{CodeJeNN} supports the following layers from the \textsc{Keras} layer API \cite{chollet_keras_2015}:
\setlength{\columnsep}{-5em}
\begin{multicols}{2}
    \begin{itemize}[itemsep=-5pt, topsep=0pt, leftmargin=*]
        \item[] \textsc{Dense}
        \item[] \textsc{Pure Activation}
        \item[] \textsc{Dropout}
        \item[] \textsc{1D, 2D, 3D Spatial Drop} \cite{tompson_efficient_2015}
        \item[] \textsc{Reshape}
        \item[] \textsc{Flatten}
        \item[] \textsc{Rescale}
        \item[] \textsc{Layer Normalization}
        \item[] \textsc{Batch Normalization}
        \item[] \textsc{Group Normalization}
        \item[] \textsc{Unit Normalization}
        \item[] \textsc{1D, 2D, 3D Convolution}
        \item[] \textsc{1D, 2D, 3D Transposed Conv} \cite{dumoulin_guide_2018, zeiler_deconvolutional_2010}
        \item[] \textsc{1D, 2D Depthwise Convolution}
        \item[] \textsc{1D, 2D Separable Convolution}
        \item[] \textsc{1D, 2D, 3D Max Pooling}
        \item[] \textsc{1D, 2D, 3D Average Pooling}
        \item[] \textsc{1D, 2D, 3D Global Max Pooling}
        \item[] \textsc{1D, 2D, 3D Global Average Pooling}
    \end{itemize}
\end{multicols}

There are also limitations to \textsc{CodeJeNN}. The primary limitation is that \textsc{CodeJeNN} only supports single-branch neural networks. The model operates on a single input tensor and does not allow multiple independent inputs; however, this input tensor may be of arbitrary dimensionality. A secondary limitation is the number of configuration options within a layer that \textsc{CodeJeNN} can support. For instance, in the Keras 3 API documentation \cite{chollet_keras_2015}, a 3D transposed convolution layer allows users to specify different regularizer options, which \textsc{CodeJeNN} does not support.
Another example includes how \textsc{CodeJeNN} handles dropout layers by skipping them entirely during code generation. Because dropout randomly zeroes activations only during training and acts as an identity operation during evaluation, \textsc{CodeJeNN} skips these layers during inference thus avoiding unnecessary operations or reading non-existent weights.
Because the code-generation pipeline supports only a subset of these configurations,
a detailed Markdown file in the GitHub repository specifies which layers \textsc{CodeJeNN} supports and to what extent, including inference options and constants.

To support development and integration, \textsc{CodeJeNN} also includes a debugging API. \textsc{Keras} has a \texttt{model.summary()} function for a trained and loaded model that provides a tabulated summary of the network architecture and layer dimensions. After code generation, \textsc{CodeJeNN} compares what is extracted from that model against what \texttt{model.summary()} provides. Additionally, even when the C\texttt{++} model generates successfully, there can still be logic/algebraic errors when performing inference. With the debug flag on, \textsc{CodeJeNN} displays all output values of each layer, which facilitates identification of errors. In practice, the majority of issues arise from user-defined custom activation functions, where the Python implementation of a custom activation does not match its C\texttt{++} counterpart.

We have also considered variadic templates as an alternative to for loops. However, such templating would quickly increase compilation time, code size, and memory usage. For instance, consider an MLP with an input sample size of $1\times 400$ and a first hidden layer of $256$ neurons. This configuration would require $102{,}400$ weights in the first layer alone, making a fully templated implementation impractical.
Consequently, this approach does not scale well to larger models due to the rapid growth in memory and compilation costs.
This issue is explained further in the \textit{Illustrative Examples} section.
For many physics applications, this limitation is acceptable because deployed models are intentionally kept small and efficient to satisfy hardware and real-time constraints.
\textsc{CodeJeNN} is intended for these use cases, where a compact model must be embedded within a larger numerical workflow. Its inlined code generation keeps the memory and computational cost of the embedded model low, although ease of integration, rather than maximum performance, is the primary aim.
If a generated model exceeds the available hardware memory, computational cost can increase substantially. In practice, however, large models are often unnecessary for accurate inference in scientific computing applications. Instead, predictive performance is frequently influenced more by factors such as data preprocessing, sampling strategies, and feature representation than by increases in model size alone \cite{zhang_multi-scale_2022}.

\section{Illustrative Examples}
We demonstrate the computational performance and implementation of \textsc{CodeJeNN} via two test cases. First, we evaluate the total wall-clock inference time and layer-wise numerical accuracy of \textsc{CodeJeNN} against Keras with an MLP and CNN. Second, we apply a neural network implementation to replace the viscosity model in a simulation of a two-dimensional subsonic splitter-plate mixing layer.
The primary objective is to demonstrate the integration of neural-network models into physics solvers, with computational benefits depending on the model and application.
These two examples are provided as detailed tutorials within the GitHub repository, which includes a comprehensive software and hardware configuration table. The respective directories contain all necessary files and instructions to fully reproduce the benchmarks presented here.

\subsection{Benchmark Test}
In our first test case, we compare the inference performance and accuracy of both an MLP and CNN created by Keras and code-generated by \textsc{CodeJeNN} using 32-bit floating-point precision. Keras requires a backend deep learning framework providing C\texttt{++} API and tensor operations, such as \textsc{TensorFlow}, \textsc{PyTorch}, or \textsc{JAX}. To ensure fair comparison, all models are tested on a single CPU core using a single thread. Since \textsc{PyTorch} and \textsc{TensorFlow} are the most widely used frameworks, only these two backends are evaluated.
By default, both \textsc{TensorFlow} and \textsc{PyTorch} perform inference in eager mode. When used independently, \textsc{TensorFlow} automatically allocates memory to an available GPU/TPU, whereas \textsc{PyTorch} defaults strictly to a CPU. When operating as a backend for Keras 3, Keras abstracts away these framework differences and standardizes device targeting. For the benchmark evaluations presented here, model initialization and inference across both backends were explicitly set to CPU configuration.

When optimizing a model for fast inference, both backends have unique functions allowing for reduced wall clock time, better memory allocation and usage, and a decrease in unnecessary computation. For \textsc{TensorFlow} as a backend, inference was optimized by freezing the entire computational graph into a single static structure with embedded weights, evaluating it within a direct graph session, and letting the runtime automatically handle internal thread allocation. For \textsc{PyTorch} as a backend, JIT optimization relies on tracing the model to map out a fixed intermediate representation, freezing model weights into immutable constants, applying inference-specific optimizations via native API calls, and disabling backpropagation tracking.

Table~\ref{tab:benchmark_arch_specs} outlines the four model architectures used in this benchmark, which vary in depth and total learnable parameters. The benchmark includes a small MLP and CNN (18,052 and 10,997 parameters, respectively), as well as a large MLP and CNN (91,836 and 843,684 parameters, respectively). These parameters are denoted as ``Params". As the layers increase in size and complexity, they showcase the range of layers \textsc{CodeJeNN} supports. Table~\ref{tab:layerwise_errors} reports the maximum absolute, mean absolute, and relative layer-wise errors between the Keras baseline and the generated C\texttt{++} models.

\begin{table}[htbp]
    \centering
    \small
    \setlength{\tabcolsep}{3pt}
    \begin{minipage}[t]{0.48\textwidth}
        \centering
        \begin{tabular}{l c w{r}{1.9cm}}
            \toprule
            \makecell[lc]{MLP\\[-3pt]Layers} & \makecell[cc]{Activation\\[-3pt]or Options} & \makecell[rc]{Output\\[-3pt]Shape}\\
            \midrule
            \multicolumn{3}{c}{Small MLP ($\approx 18$K Params)} \\
            Input & — & $(1000)$ \\
            Dense & ReLU & $(16)$ \\
            BatchNorm & — & $(16)$ \\
            Dropout & rate=0.3 & $(16)$ \\
            Dense & ReLU & $(16)$ \\
            Dense & Sigmoid & $(100)$ \\
            Output & — & $(100)$ \\
            \midrule
            \multicolumn{3}{c}{Big MLP ($\approx 92$K Params)} \\
            Input & — & $(1000)$ \\
            Dense & Linear & $(64)$ \\
            Dense & GELU & $(64)$ \\
            BatchNorm & — & $(64)$ \\
            Dense & SiLU & $(64)$ \\
            LayerNorm & — & $(64)$ \\
            Dense & ReLU & $(64)$ \\
            GroupNorm & groups=8 & $(64)$ \\
            Dense & Softplus & $(64)$ \\
            UnitNorm & — & $(64)$ \\
            Dropout & rate=0.3 & $(64)$ \\
            Dense & Tanh & $(64)$ \\
            Dense & Sigmoid & $(100)$ \\
            Output & — & $(100)$ \\
            \bottomrule
        \end{tabular}
    \end{minipage}
    \hspace{0.25em}
    \begin{minipage}[t]{0.48\textwidth}
        \centering
        \begin{tabular}{l c w{r}{1.9cm}}
            \toprule
            \makecell[lc]{CNN\\[-3pt]Layers} & \makecell[cc]{Activation\\[-3pt]or Options} & \makecell[rc]{Output\\[-3pt]Shape}\\
            \midrule
            \multicolumn{3}{c}{Small CNN ($\approx 11$K Params)} \\
            Input & — & $(10, 100, 1)$ \\
            Conv2D & Softplus & $(8, 98, 8)$ \\
            MaxPooling2D & — & $(4, 49, 8)$ \\
            Conv2D & Tanh & $(2, 47, 16)$ \\
            Conv2DTrans & ReLU & $(4, 49, 1)$ \\
            Flatten & — & $(196)$ \\
            Dense & Mish & $(32)$ \\
            Dense & Softmax & $(100)$ \\
            Output & — & $(100)$ \\
            \midrule
            \multicolumn{3}{c}{Big CNN ($\approx 843$K Params)} \\
            Input & — & $(10, 100, 1)$ \\
            Conv2D & Softplus & $(8, 98, 24)$ \\
            BatchNorm & — & $(8, 98, 24)$ \\
            MaxPooling2D & — & $(4, 49, 24)$ \\
            Conv2D & Tanh & $(2, 47, 48)$ \\
            Conv2DTrans & ReLU & $(4, 49, 32)$ \\
            SepConv2D & ELU & $(4, 49, 32)$ \\
            Flatten & — & $(6272)$ \\
            Dense & Mish & $(128)$ \\
            Dropout & rate=0.3 & $(128)$ \\
            Dense & SiLU & $(64)$ \\
            Dense & Softmax & $(100)$ \\
            Output & — & $(100)$ \\
            \bottomrule
        \end{tabular}
    \end{minipage}
    \caption{Model architecture of all benchmark test models. This table includes the output shape of each layer and any activation functions or options appended to that layer. For example, a dropout layer does not have an activation function; instead, the dropout rate is listed in the "Activation" column. For convolutional and pooling layers, output shapes are formatted as a 3-tuple $(H, W, C)$, representing the spatial height, spatial width, and channel (or filter) depth of the feature map. "Params" denotes the total approximate learnable parameters.}
    \label{tab:benchmark_arch_specs}
\end{table}

\begin{table}[htbp]
    \centering
    \small
    \setlength{\tabcolsep}{2.5pt}
    \begin{minipage}[t]{0.48\textwidth}
        \centering
        \begin{tabular}{lccc}
            \toprule
            \makecell[lc]{MLP\\[-3pt]Layers} & \makecell[cc]{Max\\[-3pt]Abs\\[-3pt]Error} & \makecell[cc]{Mean\\[-3pt]Abs\\[-3pt]Error} & \makecell[cc]{Rel\\[-3pt]Error} \\
            \midrule
            \multicolumn{4}{c}{Small MLP ($\approx 18$K Params)} \\
            1 (Dense) & 3.89E-4 & 5.28E-5 & 5.81E-7 \\
            2 (BatchNorm) & 4.00E-4 & 5.32E-5 & 5.57E-7 \\
            3 (Dropout) & — & — & — \\
            4 (Dense) & 1.98E-4 & 4.28E-5 & 1.05E-6 \\
            5 (Dense) & 8.49E-6 & 3.74E-7 & 2.05E-6 \\
            \midrule
            \multicolumn{4}{c}{Big MLP ($\approx 92$K Params)} \\
            1 (Dense) & 1.36E-3 & 2.43E-4 & 4.26E-7 \\
            2 (Dense) & 9.15E-4 & 1.49E-4 & 4.17E-7 \\
            3 (BatchNorm) & 9.77E-4 & 1.86E-4 & 4.22E-7 \\
            4 (Dense) & 7.40E-4 & 1.59E-4 & 4.27E-7 \\
            5 (LayerNorm) & 1.65E-6 & 2.80E-7 & 4.70E-7 \\
            6 (Dense) & 1.37E-6 & 2.28E-7 & 6.29E-7 \\
            7 (GroupNorm) & 3.91E-6 & 6.40E-7 & 1.03E-6 \\
            8 (Dense) & 1.79E-6 & 4.48E-7 & 6.54E-7 \\
            9 (UnitNorm) & 2.70E-7 & 6.04E-8 & 6.49E-7 \\
            10 (Dropout) & — & — & — \\
            11 (Dense) & 2.16E-7 & 7.02E-8 & 9.35E-7 \\
            12 (Dense) & 1.17E-7 & 3.24E-8 & 8.43E-08 \\
            \bottomrule
        \end{tabular}
    \end{minipage}%
    \hspace{0.25em}
    \begin{minipage}[t]{0.49\textwidth}
        \centering
        \begin{tabular}{lccc}
            \toprule
            \makecell[lc]{CNN\\[-3pt]Layers} & \makecell[cc]{Max\\[-3pt]Abs\\[-3pt]Error} & \makecell[cc]{Mean\\[-3pt]Abs\\[-3pt]Error} & \makecell[cc]{Rel\\[-3pt]Error} \\
            \midrule
            \multicolumn{4}{c}{Small CNN ($\approx 11$K Params)} \\
            1 (Conv2D) & 5.40E-5 & 2.75E-6 & 6.46E-8 \\
            2 (MaxPool2D) & 5.40E-5 & 2.95E-6 & 6.34E-8 \\
            3 (Conv2D) & 1.31E-5 & 8.22E-8 & 7.14E-7 \\
            4 (Conv2DTrans) & 2.00E-6 & 1.37E-7 & 5.49E-7 \\
            5 (Flatten) & — & — & — \\
            6 (Dense) & 8.37E-7 & 2.13E-7 & 7.84E-7 \\
            7 (Dense) & 7.80E-9 & 3.43E-9 & 3.75E-7 \\
            \midrule
            \multicolumn{4}{c}{Big CNN ($\approx 843$K Params)} \\
            1 (Conv2D) & 6.10E-5 & 1.43E-6 & 4.56E-8 \\
            2 (BatchNorm) & 1.83E-4 & 5.88E-6 & 7.91E-8 \\
            3 (MaxPool2D) & 1.83E-4 & 6.65E-6 & 8.15E-8 \\
            4 (Conv2D) & 6.27E-6 & 9.88E-9 & 1.63E-7 \\
            5 (Conv2DTrans) & 1.13E-6 & 7.34E-8 & 3.36E-7 \\
            6 (SepConv2D) & 2.01E-7 & 2.36E-8 & 3.96E-7 \\
            7 (Flatten) & — & — & — \\
            8 (Dense) & 7.00E-7 & 1.06E-7 & 1.36E-6 \\
            9 (Dropout) & — & — & — \\
            10 (Dense) & 3.15E-7 & 7.66E-8 & 1.30E-6 \\
            11 (Dense) & 7.40E-9 & 3.70E-9 & 3.86E-7 \\
            \bottomrule
        \end{tabular}
    \end{minipage}
    \caption{This table shows the layer-wise numerical errors for each benchmark test model. All values for each layer within a model are printed for both a model created by Keras and code-generated by \textsc{CodeJeNN}. Since the C\texttt{++} model is generated from Keras, Keras serves as the reference values to validate against. The table reports maximum absolute error, mean absolute error, and relative error per layer. Layers such as Flatten and Dropout do not output any values and are denoted with "-". This type of error calculation is also available as a debugging API within \textsc{CodeJeNN} to help with identification of discrepancies in code-generated models. 32-bit precision was used to train, generate, and perform inference with all models.}
    \label{tab:layerwise_errors}
\end{table}

\begin{table}[h]
    \centering
    \small
    \setlength{\tabcolsep}{2.5pt}
    \begin{tabular}{@{} l ccc ccc ccc ccc @{}}
        \toprule
        & \multicolumn{3}{c}{\makecell{Small MLP\\($\approx 18$K Params)}}
        & \multicolumn{3}{c}{\makecell{Big MLP\\($\approx 92$K Params)}}
        & \multicolumn{3}{c}{\makecell{Small CNN\\($\approx 11$K Params)}}
        & \multicolumn{3}{c}{\makecell{Big CNN\\($\approx 843$K Params)}} \\
        \cmidrule(lr){2-4} \cmidrule(lr){5-7} \cmidrule(lr){8-10} \cmidrule(l){11-13}
        \smash{\makecell[lb]{Keras Deep\\Learning\\Backend}}
        & Avg & Std & Med & Avg & Std & Med & Avg & Std & Med & Avg & Std & Med \\
        \midrule
        \makecell[lc]{Tensorflow\\[-3pt]Eager}
        & 9.43 & 9.72E-2 & 9.42
        & 30.47 & 2.97E-1 & 30.49
        & 15.81 & 2.94E-1 & 15.70
        & 26.33 & 3.99E-1 & 26.22 \\
        \makecell[lc]{PyTorch\\[-3pt]Eager}
        & 8.46 & 7.47E-2 & 8.44
        & 19.37 & 2.76E-1 & 19.31
        & 11.66 & 1.26E-1 & 11.62
        & 22.10 & 1.20E-1 & 22.09 \\
        \makecell[lc]{Tensorflow\\[-3pt]Optimized}
        & 0.35 & 3.59E-3 & 0.35
        & 0.43 & 5.09E-3 & 0.43
        & 0.60 & 3.31E-3 & 0.60
        & 2.59 & 7.08E-2 & 2.57 \\
        \makecell[lc]{PyTorch\\[-3pt]Optimized}
        & 0.11 & 4.83E-3 & 0.11
        & 0.37 & 5.37E-3 & 0.37
        & 0.77 & 1.47E-2 & 0.77
        & 4.05 & 5.51E-2 & 4.05 \\[8pt]
        \textbf{\textsc{CodeJeNN}}
        & 0.01 & 6.39E-4 & 0.01
        & 0.05 & 6.15E-4 & 0.05
        & 0.32 & 1.24E-3 & 0.32
        & 2.06 & 5.09E-3 & 2.06 \\
        \bottomrule
    \end{tabular}
    \caption{MLP and CNN inference comparisons. The inference times represent the total wall-clock time (seconds) to compute the inference of 10,000 random data samples sequentially for an MLP and CNN of various sizes, created by Keras and \textsc{CodeJeNN}. We test the Keras backend inference speed of \textsc{PyTorch} and \textsc{TensorFlow} in both eager mode and when fully optimized. Each inference case is tested 30 times to ensure precision. "Params" stands for the total approximate learnable parameters of that model or degrees of freedom. All models were trained, generated, and executed in 32-bit precision.}
    \label{tab:mlp_inference}
\end{table}

Qualitatively, Table~\ref{tab:layerwise_errors} shows an accumulation of errors spanning all layers, as each layer carries the numerical deviation of preceding layers alongside its own computation. Because the primary objective of \textsc{CodeJeNN} is to embed trained models into established physics solvers, this accumulation is insignificant so long as the final output remains within acceptable error tolerances. As shown in Table~\ref{tab:layerwise_errors}, the final layer across all four benchmark models produces smaller absolute errors compared to the preceding hidden layers. This occurs because while the relative error remains consistently bounded near 32-bit floating-point precision limits ($10^{-6}$ to $10^{-8}$) across all layers, the smaller magnitudes and squashing activations at the final output layer naturally contract the absolute error down to $10^{-7}$ to $10^{-9}$.


The limitation of memory described earlier is explained here. From our experience, because every weight/bias is written as a compile-time constant in the generated header file, the time and memory for compilation increase with the number of learnable parameters. The largest benchmark model tested ($\approx 843$K parameters) compiles within seconds, whereas significantly larger models can exhaust the compiler memory. A second effect appears during runtime. Once the weights/biases and layer outputs of a model no longer fit within the processor's cache, inference becomes limited by memory access rather than arithmetic throughput, and the benefit of inlined code diminishes. As shown in Table~\ref{tab:mlp_inference}, the performance margin between \textsc{CodeJeNN} and optimized Keras narrows as model capacity increases.

\subsection{Inference Demonstration on a Splitter Plate}
We embedded a neural network to replace viscosity (computed with Wilke's mixing rule \cite{wilke_viscosity_1950} assuming mixture-averaged transport) in a hydrogen-air splitter-plate mixing layer simulation using \textsc{OpenFOAM} \cite{weller_tensorial_1998} and \textsc{DetonationFOAM} \cite{sun_detonationfoam_2023}. While viscosity evaluation is typically a minor cost in multicomponent flow solvers, the purpose here is to demonstrate that neural networks can be robustly embedded within a CFD workflow using \textsc{CodeJeNN}.

In the baseline \textsc{DetonationFOAM} case, the mixture-averaged dynamic viscosity is calculated from species-dependent viscosity fits and mole fractions as
\begin{equation}
    \mu_{\mathrm{mix}}
    =
    \sum_{i=1}^{N}
    \frac{X_i \mu_i}
         {\sum_{j=1}^{N} X_j \Phi_{ij}}
    \label{eq:wilke}
    ,
\end{equation}
where $X_i$ is the mole fraction of species $i$, and the interaction coefficient is
\begin{equation}
    \Phi_{ij} =
    \frac{1}{\sqrt{8}}
    \left(1 + \frac{M_i}{M_j}\right)^{-1/2}
    \left[
        1 +
        \left(\frac{\mu_i}{\mu_j}\right)^{1/2}
        \left(\frac{M_j}{M_i}\right)^{1/4}
    \right]^2,
    \label{eq:phi}
\end{equation}
with $M_i$ denoting the molar mass of species $i$. The species viscosity is evaluated using a log-polynomial fit in temperature:
\begin{equation}
    \mu_i =
    0.1\exp\!\left(
        A_{1,i}
        + A_{2,i}\ln T
        + A_{3,i}(\ln T)^2
        + A_{4,i}(\ln T)^3
    \right)
    \quad [\text{Pa$\cdot$ s}].
    \label{eq:muspecies}
\end{equation}
The domain is $15\,\text{mm}\times5\,\text{mm}\times0.2\,\text{mm}$. Two streams enter from the left and are separated at the centerplane at $y=2.5\,\text{mm}$:
\begin{itemize}
    \item \textbf{H$_2$ stream}: $M=0.3$, $T=350\,\text{K}$, $p=101{,}325\,\text{Pa}$,
    $U\approx427\,\text{m/s}$.
    \item \textbf{Air stream}: $M=0.3$, $T=1500\,\text{K}$, $p=101{,}325\,\text{Pa}$,
    $U\approx226\,\text{m/s}$.
\end{itemize}

The mesh consists of two blocks per stream with $190 \times 60 \times 1$ cells each. Streamwise grading ranges from $20\,\mu\text{m}$ at the inlet to $200\,\mu\text{m}$ at the outlet, with wall-normal grading symmetric about the centerplane ($10\,\mu\text{m}$ at the shear interface and $\approx 100\,\mu\text{m}$ at boundaries). The top, bottom, and right boundaries are prescribed as zero-gradient supersonic outflow.

\begin{figure}[H]
    \includegraphics[width=\textwidth]{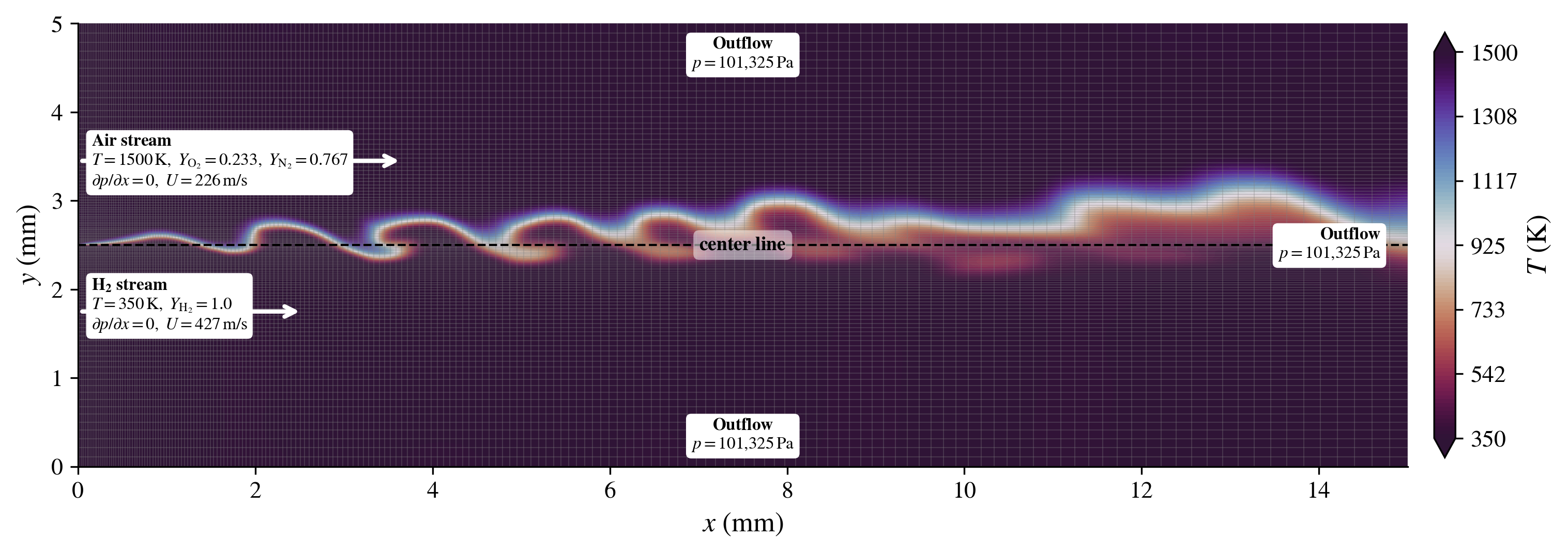}
    \caption{Temperature and flow domain}
    \label{fig:cfd_domain}
\end{figure}

Chemistry in \textsc{DetonationFOAM} is disabled to reduce computational cost. A quasi-steady flow field after five flow-through times ($5 \times L/U_{x,air}\approx 3.32 \times 10^{-4}$~s) is shown in Figure~\ref{fig:cfd_domain}, where unsteady mixing layer structures grow with downstream distance. Wilke's model is replaced with an MLP generated by \textsc{CodeJeNN} that maps thermochemical state directly to viscosity: $\left(Y_{\mathrm{H}_2},Y_{\mathrm{O}_2},Y_{\mathrm{N}_2},T\right)\rightarrow\mu_{\text{mix}}$.

The MLP is trained on $10^4$ random states $(Y_{H_2},Y_{O_2},Y_{N_2},T)$ generated from a small OpenFOAM utility, \texttt{generateData}, available in the \textsc{CodeJeNN} repository, which uses the viscosity from \texttt{detonationFOAM} as the training target. For 60\% of samples, the mass fractions are obtained from a symmetric Dirichlet distribution and temperature is obtained from a uniform distribution as $T\sim\mathcal{U}(200,3000)$\,K. To increase coverage of states that are observed in the domain, the remaining 40\% are obtained along the physical mixing line as
\begin{equation}
    \mathbf{Y}(\alpha) = \alpha\,\mathbf{Y}_{\text{fuel}}
                       + (1-\alpha)\,\mathbf{Y}_{\text{ox}}, \quad
    \alpha\sim\mathcal{U}(0,1),
\end{equation}
where $\mathbf{Y}_{\text{fuel}}=(1,0,0)$ (pure H$_2$) and
$\mathbf{Y}_{\text{ox}}=(0,\,0.233,\,0.767)$ (air), and
\begin{equation}
    T(\alpha)=\alpha T_{\text{fuel}}+(1-\alpha)T_{\text{ox}}+\delta T,\quad
    \delta T\sim\mathcal{U}(-200,200)\text{ K},
\end{equation}
where $T_{\text{fuel}}=350$\,K and
$T_{\text{ox}}=1500$\,K.

The MLP consists of 3 hidden layers with 12 neurons each (385 parameters total) and Tanh activation. Normalized inputs and de-normalized outputs are embedded directly in C\texttt{++} headers within a modified \textsc{DetonationFOAM} solver, requiring no external files at runtime. The network is verified against the Wilke model over 2,000 randomly sampled states independent of the $10^4$ training samples. The model is trained and performs inference using 64-bit precision. The relative error for each sample is computed as
\begin{equation}
    \epsilon = {100 \times} \frac{|\mu_{\text{Wilke}} - \mu_{\text{NN}}|}{\mu_{\text{Wilke}}},
    \label{error}
\end{equation}
yielding a mean error of 0.12\% and a maximum error of 8.12\%.

\begin{figure}[H]
    \includegraphics[width=\textwidth]{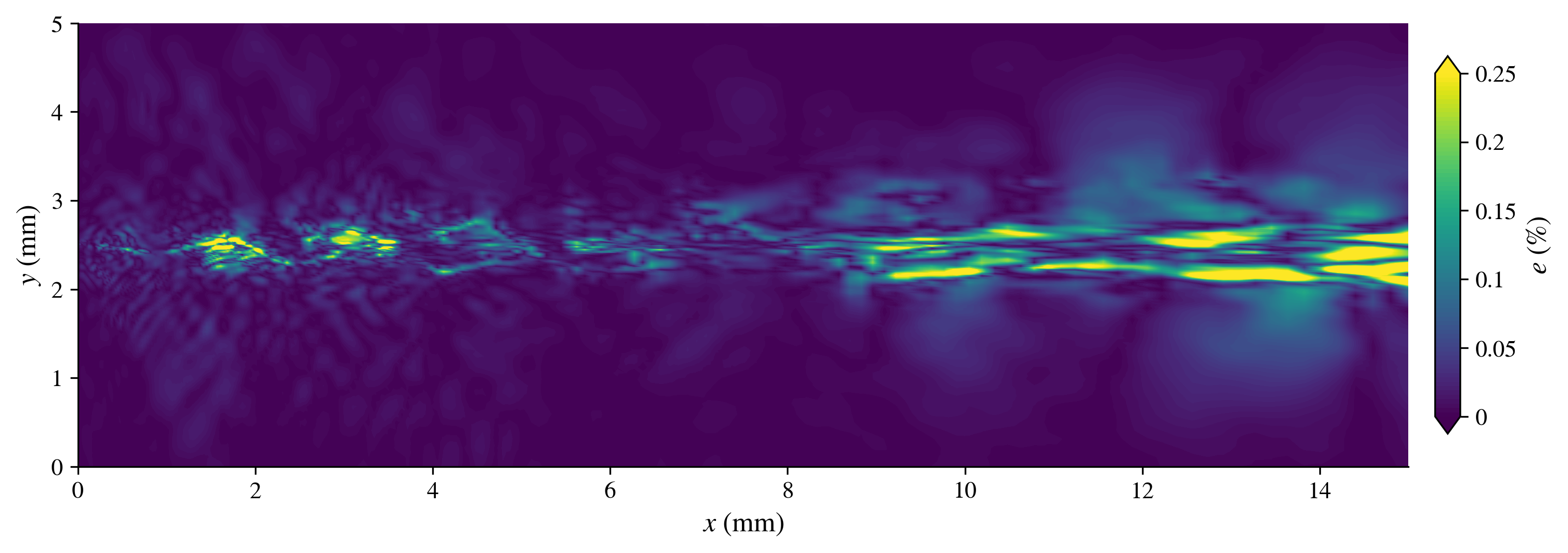}
    \caption{Error in $x$-velocity after one flow-through time}
    \label{fig:u_error}
\end{figure}

Two simulations are run for five more flow-through times: one using the native Wilke model and another using the generated MLP. Figure~\ref{fig:u_error} presents the percent error for the $x$-velocity field, computed as $100\times\left|u_{x, Wilke}-u_{x, NN}\right|/u_{x, Wilke}$.
Table \ref{tab:field_errors} represents the error metrics for Figure~\ref{fig:cfd_domain} and Figure~\ref{fig:u_error}. The neural network introduces minimal error: the relative $L_2$ error across the domain is 0.065\% for $x$-velocity and 0.11\% for temperature, with mean absolute errors of 0.100~m/s and 0.406~K, respectively. Localized maximum absolute errors reach 3.74~m/s for $x$-velocity and 9.51~K for temperature strictly within the high-gradient mixing layer immediately downstream of the splitter plate.
The absence of large-scale errors elsewhere indicates that the neural-network viscosity model preserves overall flow evolution. These results demonstrate that CodeJeNN-generated networks can be embedded in CFD solvers as surrogates for constitutive models while maintaining close agreement with reference solutions.

\begin{table}[htbp]
    \centering
    \small
    \begin{tabular}{l c c c}
        \toprule
        Field & Rel $L_2$ Error & Mean Abs Error & Max Abs Error \\
        \midrule
        Temperature & 0.11\% & 0.406 K & 9.51 K \\
        $x$-velocity & 0.065\% & 0.100 m/s & 3.74 m/s \\
        \bottomrule
    \end{tabular}
    \caption{Error metrics of the $T$ and $u$ fields between the reference solver and the neural-network-embedded solver.}
    \label{tab:field_errors}
\end{table}

\section{Impact}
\textsc{CodeJeNN} allows researchers to integrate neural networks into their computational physics codebases with minimal additional infrastructure. This product serves as a way to build inline infrastructure with only the prerequisite of a basic understanding of how neural networks perform inference. The key feature of \textsc{CodeJeNN} is its ability to produce raw source code where everything is inlined and contained within a single header file. This circumvents the need to export tensor operations from high-performance ML libraries, preventing dependency conflicts and avoiding thread oversubscription in physics codes that already utilize all available hardware.

Researchers who are not ML specialists increasingly rely on LLMs to integrate neural networks into their computational physics codes. While this lowers the barrier to entry, it introduces a recurring cost that \textsc{CodeJeNN} is designed to avoid.
Integrating neural networks into an established codebase using an LLM may require repeatedly supplying project-specific context and regenerating infrastructure whenever architectures change. Prior work has shown that software-engineering workflows can consume substantial token budgets, with context processing accounting for a large fraction of total token usage \cite{salim_tokenomics_2026}. By automatically generating self-contained inference code, \textsc{CodeJeNN} reduces reliance on repeated LLM-assisted infrastructure generation and the associated token costs. Additionally, it has been shown that LLM reasoning often contains redundant tokens that increase inference costs without providing proportional gains in performance \cite{han_token-budget-aware_2025}. \textsc{CodeJeNN} addresses this challenge by eliminating the need to repeatedly generate infrastructure code through an LLM, instead providing an easy-to-call function that computational physics practitioners can integrate directly into their codes.

Finally, \textsc{CodeJeNN} is designed to be extensible and repeatable. If a particular architecture does not exist within \textsc{CodeJeNN}, a user can easily leverage an LLM to add the required logic with substantially fewer tokens. This enables a non-invasive workflow for generating new models while minimizing token consumption and user effort.

\section{Conclusion}
\textsc{CodeJeNN}, a C\texttt{++} neural network generation library, has been developed and demonstrated. This library is compatible with Keras, a commonly used deep learning API in computational physics, and enables users to embed neural networks into their codebases. The package does so by providing a self-contained C\texttt{++} rendition of the Keras model. This allows the user to integrate the model directly into their solver and, if needed, optimize the raw source code further. We have demonstrated the capabilities of \textsc{CodeJeNN} by evaluating its inference speed and accuracy and replacing the viscosity model in a subsonic splitter-plate mixing layer simulation.

Future work will integrate \textsc{Spektral} \cite{grattarola_graph_2020}, which is the Graph Neural Network (GNN) API for Keras that includes commonly used GNN layers such as message passing layers \cite{gilmer_neural_2017} and attention-based graphing layers \cite{thekumparampil_attention-based_2018}. Additional activation functions will also be added. Furthermore, we will explore accelerating the inference of generated models by adding features to export functions from low-level high-performance libraries such as \textsc{Eigen} \cite{guennebaud_eigen_2010-1}. This path may also include using different data structures for model hyperparameters, allowing users to homogenize the neural network within their codebase. \textsc{CodeJeNN} is accessible at \url{https://github.com/jarcities/codejenn}.

\section{Conflict of Interest}


We wish to confirm that there are no known conflicts of interest associated with this publication and there has been no significant financial support for this work that could have influenced its outcome.

\section*{Acknowledgements}

The authors wish to acknowledge Dr. Eric Marineau and Dr. Jonathan Sosa of the Hypersonic Aerothermodynamics, High-Speed Propulsion and Materials Program of the Office of Naval Research Code 35 for directly supporting this work.

\section*{Declaration of generative AI and AI-assisted technologies in the manuscript preparation process}
During the preparation of this work the author(s) used Claude from Anthropic in order to verify CFD test cases after infrastructure was already implemented.
After using this tool/service, the author(s) reviewed and edited the content as needed and take(s) full responsibility for the content of the published article.

\bibliographystyle{elsarticle-num}
\bibliography{References}

\end{document}